# Пикосекундное переключение высоковольтных обратносмещенных $p^+$-$n$-$n^+$-структур в проводящее состояние при импульсном освещении.

А. С. Кюрегян

Всероссийский Электротехнический институт им. В. И. Ленина, 111250, Москва, Россия
E-mail: ask@vei.ru

Построена аналитическая теория пикосекундного переключение высоковольтных обратносмещенных $p^+$-$n$-$n^+$-структур в проводящее состояние при воздействии импульсного освещения и проведено численное моделирование этого процесса. Объединение результатов теории и моделирования позволило получить простые соотношения между параметрами структуры, светового импульса, внешней цепи и основными характеристиками процесса – амплитудой импульса тока активной нагрузки и длительностью процесса коммутации.

## 1. Введение

Изучение мощных субнаносекундных полупроводниковых коммутаторов, управляемых короткими импульсами оптического излучения, началось более 40 лет назад [1-2] и продолжается до сих пор (см., например, [3-5]). В большинстве экспериментов в качестве полупроводникового элемента использовались однородные высокоомные кристаллы или пленки с омическими контактами (то есть по сути дела – фотосопротивления), встроенные в разрыв микрополосковой линии. Предельная простота конструкции оказалась весьма привлекательной для многих исследователей, продемонстрировавших уникальные возможности таких приборов. Однако недостаточно высокое темновое сопротивление сильно ограничивает предельную напряженность поля при постоянном смещении и, вследствие этого, требует применения импульсных источников питания. Этого недостатка фотосопротивлений лишены высоковольтные фотодиоды, в которых и при постоянном смещении легко достижимы напряженности поля, близкие к пробивным. Первые эксперименты с фотодиодными коммутаторами были описаны в статьях [6-8], но дальнейшего развития эти работы не получили. Вероятной причиной этого является то, что теория процесса коммутации высоковольтных фотодиодов, необходимая для их проектирования, по сути дела, отсутствует. Единственное известное нам исследование [9] посвящено лишь заключительной стадии процесса коммутации, во время которой практически все напряжение, блокируемое до подачи управляющего импульса, уже перераспределилось на нагрузку. Начальная же стадия, определяющая быстродействие коммутатора, насколько нам известно, не изучалась. Этому вопросу и посвящена настоящая работа.

## 2. Модель фотодиода

В качестве модели для анализа процесса коммутации мы используем кремниевую $p^+$-$n$-$n^+$-структуру площадью $S_d$, схематически изображенную на рис. 1а. На торцевые поверхности структуры наложены два металлических электрода: сплошной со стороны $n^+$-слоя и кольцевой со стороны $p^+$-слоя. Центральная область поверхности $p^+$-слоя площадью $S_0$ открыта для проникновения управляющего импульса оптического излучения внутрь структуры. Будем считать, что

- $n$-слой однородно легирован донорами с концентрацией $N_d$ ( рис. 1б), достаточно малой для того, чтобы при начальном обратном смещении структуры $U_0$ истощенная область с сильным электрическим полем $E$ занимает всю базу (рис. 1в);

- концентрации доноров $N_d^+$ и акцепторов $N_a^+$ в тонких $n^+$- и $p^+$-слоях столь велики, что их можно рассматривать как продолжение металлических электродов[1];
- толщина $n$-слоя $d$ много меньше его поперечных размеров;
- управляющее оптическое излучение падает перпендикулярно поверхности окна и освещает его однородно.

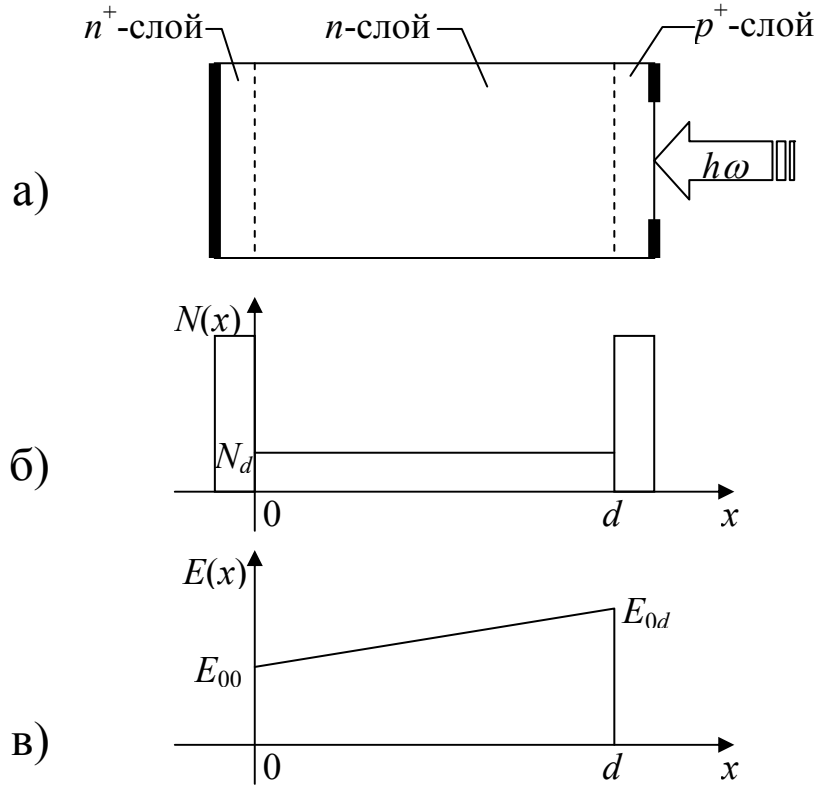

Рис. 1. Схематическое изображение конструкции фотодиодного коммутатора (а), распределение легирующих примесей (б) и электрического поля до начала коммутации (в).

Эквивалентная схема коммутатора, включенного последовательно с источником постоянного обратного смещения $U_0$ и активной нагрузкой $R_L$ изображена на рис. 2. До подачи управляющего оптического импульса сопротивление фотодиода на много порядков больше $R_L$, так что емкость $C_d = \varepsilon S_d / d$ ($\varepsilon$ - диэлектрическая проницаемость полупроводника) заряжена до напряжения $U_0$, а ток нагрузки близок к нулю. После начала освещения фотодиода его сопротивление резко уменьшается, емкость $C$ начинают разряжаться, в результате чего все большая часть внешнего смещения падает на нагрузке, через которую начинает протекать возрастающий со временем ток $I_L(t)$. В этом по сути дела и заключается процесс коммутации. Для его описания на качественном уровне можно считать, что в результате освещения емкость фотодиода **мгновенно** шунтируется проводящим каналом с **постоянным линейным** сопротивлением $R_d$. Тогда ток нагрузки начнет возрастать по закону

---

[1] Наибольшая ошибка при этом связана с пренебрежением продольным сопротивлением $p^+$-слоя в окне фотодиода. Для круглого окна разность потенциалов между его центром и краем, обусловленная протеканием тока с плотностью $J$, как легко убедиться, равна $\rho_s J S_0 / 4\pi$, где $\rho_s$ - поверхностное сопротивление $p^+$-слоя. Поэтому $R_s \approx \rho_s / 4\pi = 0{,}1$-$0{,}5$ Ом для типичных значений $\rho_s$, тогда как обычное значение $R_L$ на 2-3 порядка больше.

$$I_L(t) = \frac{U_0}{R_L + R_d}\left(1 - e^{-t/\tau}\right) \quad (1)$$

где $\tau = C_d R_L R_d / (R_L + R_d)$. Однако при расчете ампер-секундной характеристики реального коммутатора необходимо учитывать, что длительность $t_0$ импульса света **конечна**, а эффективное сопротивление $R_d$ освещенной части фотодиода **нелинейно и непостоянно**.

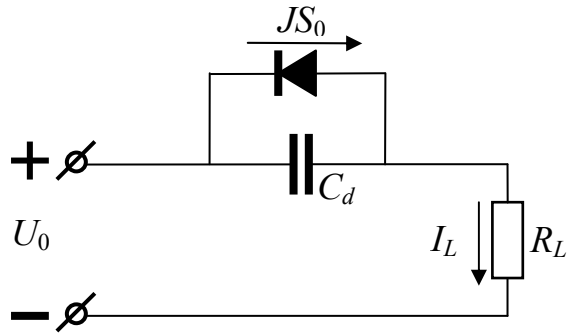

Рис. 2. Эквивалентная схема, использованная для описания процесса коммутации.
$J$ - полная плотность тока в освещенной области фотодиода площадью $S_0$.

Для решения этой задачи мы воспользуемся обобщенным соотношением Шокли-Рамо [10-12], согласно которой в нашем случае ток через нагрузку равен

$$I_L = S_0 \bar{j} + C_d \frac{dU}{dt}, \quad (2)$$

где $U = (U_0 - I_L R_L)$ - мгновенное падение напряжения на диоде,

$$\bar{j} = \frac{1}{d}\int_0^d j\, dx, \quad (3)$$

$j = j_n + j_p$ - плотность тока электронов и дырок в освещенной области $n$-слоя. Соотношение (2) можно переписать в виде простого уравнения[2]

$$\tau_d \frac{dI_L}{dt} + I_L = S_0 \bar{j} \quad (4)$$

для искомой ампер-секундной характеристики $I_L(t)$, решение которого с очевидным начальным условием $I_L(0) = 0$ имеет вид

$$I_L(\theta) = S_0 \int_0^\theta \bar{j}(\theta') \exp(\theta' - \theta) d\theta', \quad (5)$$

где введено безразмерное время $\theta = t/\tau_d$, $\tau_d = R_L C_d$. Формула (5) описывает процесс коммутации в общем виде и сводит задачу к вычислению $\bar{j}(t)$. Ее можно найти, зная распределения электронов, дырок и электрического поля в освещенной области $n$-слоя.

---

[2] Оно представляет собой обобщение уравнения (2) из работы [13].

### 3. Аналитическая теория процесса коммутации.

При сделанных выше предположениях распределения рожденных светом носителей заряда в истощенном слое описываются одномерными уравнениями непрерывности

$$\frac{dn}{dt} - \frac{1}{q}\frac{dj_n}{dx} = G, \qquad \frac{dp}{dt} + \frac{1}{q}\frac{dj_p}{dx} = G \qquad (6)$$

с граничными и начальными условиями

$$n(d,t) = p(0,t) = 0, \qquad (7)$$

$$n(x,0) = p(x,0) = 0 \qquad (8)$$

В дрейфовом приближении плотности токов проводимости $j_{n,p}$ равны

$$j_n = qnv_n, \qquad j_p = qpv_p, \qquad (9)$$

где $q$ - заряд электрона, $n$ и $p$ - концентрации электронов и дырок, $v_n$ и $v_p$ - их дрейфовые скорости, $G$ - скорость генерации пар в истощенном слое.

Аналитическое решение этой задачи можно получить, предполагая, что во всей базе напряженность поля $E$ превосходит пороговое значение $E_s \approx 15$ кВ/см, выше которого дрейфовые скорости перестают зависеть от $E$. Для практически наиболее интересного случая, когда начальное смещение $U_0$ близко к напряжению пробоя структуры $U_B$, а поле $E_{0d}$ – к пробивному значению $E_B \approx 150$ кВ/см, это предположение оправдано в течение почти всего процесса коммутации в силу неравенства $E_B \gg E_s$. Тогда уравнения непрерывности можно решать независимо от уравнения Пуассона и внешней цепи. Результат удобно представить в виде формул

$$p(x,t) = \int_{T_p}^{t} G[x - v_p(t-t'), t']dt', \qquad n(x,t) = \int_{T_n}^{t} G[x + v_n(t-t'), t']dt', \qquad (10)$$

справедливость которых нетрудно проверить простой подстановкой в (6). Нижние пределы интегрирования в (10), определяемые равенствами

$$T_p = \max(0, t - x/v_p), \qquad T_n = \max[0, t - (d-x)/v_n],$$

появились вследствие граничных условий (7) и условия $G(x,t) = 0$ при $t < 0$.

Если оптическое излучение представляет собой плоскую волну, падающую на фотодиод вдоль оси $x$, то скорость генерации пар можно представить в виде

$$G(x,t) = G_0 f(t) e^{\kappa x}, \qquad (11)$$

где

$$G_o = P_{ph} \frac{\kappa(1 - R_{ph})}{\hbar \omega S_o} e^{-\kappa(d + x_j)},$$

$P_{ph}$ - пиковая мощность импульса, $\hbar\omega$ - энергия кванта, $\kappa$ - коэффициент поглощения света в полупроводнике, $R_{ph}$ - коэффициент отражения света от поверхности окна фотодиода, $x_j$ - толщина $p^+$-слоя, а $f(t)$ – функция, описывающая форму управляющего импульса. При написании (11) мы пренебрегали отражением света от тылового контакта, возможной (например, вследствие эффекта Франца-Келдыша) зависимостью $\kappa$ от координаты и лавинным размножением носителей заряда. В дальнейшем для определенности будем считать импульс излучения прямоугольным: $f(t) = 1$ при $0 < t < t_0$ и $f(t) = 0$ при других $t$. Подставляя (11) в (10), можно получить следующие распределения электронов и дырок в базе при $0 < t < t_0$

$$p(x,t) = \frac{G_0}{\kappa v_p} \begin{cases} \left(e^{\kappa x}-1\right) & \text{при} \quad 0 < x < v_p t \\ \left(1-e^{-\kappa v_p t}\right)e^{\kappa x} & \text{при} \quad v_p t < x < d \end{cases}, \quad (12)$$

$$n(x,t) = \frac{G_o}{\kappa v_n} \begin{cases} \left(e^{\kappa d}-e^{\kappa x}\right) & \text{при} \quad d-v_n t < x < d \\ \left(e^{\kappa v_n t}-1\right)e^{\kappa x} & \text{при} \quad 0 < x < d-v_n t \end{cases}, \quad (13)$$

и при $t > t_0$

$$p(x,t) = \frac{G_0}{\kappa v_p} \begin{cases} 0 & \text{при} \quad 0 < x < v_p(t-t_0) \\ \left[e^{\kappa x - \kappa v_p(t-t_0)}-1\right] & \text{при} \quad v_p(t-t_0) < x < v_p t, \\ \left(e^{\kappa v_p t_0}-1\right)e^{\kappa x - \kappa v_p t} & \text{при} \quad v_p t < x < d \end{cases} \quad (14)$$

$$n(x,t) = \frac{G_0}{\kappa v_n} \begin{cases} 0 & \text{при} \quad d-v_n(t-t_0) < x < d \\ \left[e^{\kappa d}-e^{-\kappa x + \kappa v_n(t-t_0)}\right] & \text{при} \quad d-v_n t < x < d-v_n(t-t_0). \\ \left(1-e^{-\kappa v_n t_0}\right)e^{\kappa x + \kappa v_n t} & \text{при} \quad 0 < x < d-v_n t \end{cases} \quad (15)$$

Пример таких распределений приведен на Рис. 3. Они, как и все последующие количественные результаты и оценки, иллюстрирующие процесс коммутации, получены для кремниевых диодов при следующих значениях параметров: $d = 0.5$ мм, $N_d = 10^{13}$ см$^{-3}$, $S_d = 10$ мм$^2$, $S_0 = 6$ мм$^2$, $U_0 = 6$ кВ, $R_L = 50$ Ом, $\tau_d = 106$ пс, $\hbar\omega = 1.16$ эВ, $\kappa = 7$ см$^{-1}$, $R_{ph} = 0.5$.

Подстановка (9),(12)-(15) в (3) дает среднюю плотность тока проводимости:

$$\bar{j}(t) = \frac{qG_0}{\kappa^2 d} \begin{cases} 1-\kappa v_p t - e^{\kappa v_n t} + \left(1+\kappa v_n t - e^{-\kappa v_p t}\right)e^{\kappa d} & \text{при} \quad 0 < t < t_0 \\ \kappa v_n t_0 e^{\kappa d} - \left(1-e^{-\kappa v_n t_0}\right)e^{\kappa v_n t} - \kappa v_p t_0 + \left(e^{\kappa v_p t_0}-1\right)e^{\kappa d - \kappa v_p t} & \text{при} \quad t_0 < t \end{cases} \quad (16)$$

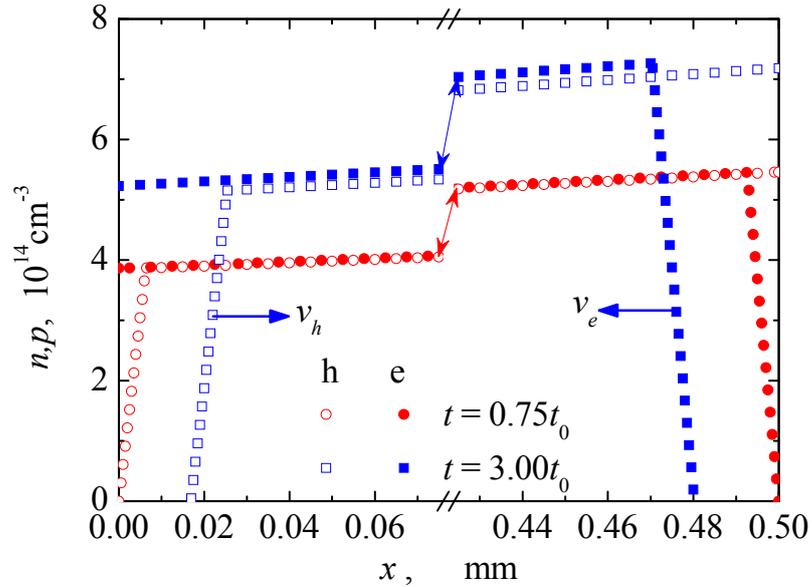

Рис. 3. Распределения электронов (темные символы) и дырок (светлые символы) в n-слое фотодиода в моменты времени $t = 0.75 t_0$ (квадраты) и $t = 3.0 t_0$ (кружки) после начала освещения импульсом света с энергией $W_{ph} = 1.95$ мкДж и длительностью $t_0 = 100$ пс.

Наибольший практический интерес представляет случай быстрой коммутации, происходящей за время $t_c$, много меньшее времени $t_{np} = d/(v_n + v_p)$ пролета электронов и дырок через базу под действием короткого ($t_0 \ll t_{np}$) импульса не очень сильно поглощаемого излучения ($\kappa d \leq 1$). В этом случае можно разложить экспоненты в формуле (16) в ряд по малости величин $\kappa v_{n,p} t$, что дает

$$\bar{j}(t) = \frac{W_{ph}}{S_0 U_\omega t_{np}} \begin{cases} \theta\left(1 - \mu \dfrac{\theta}{2}\right) & \text{при} \quad \theta < \theta_0 \\ \theta_0 \left[1 - \mu\left(\theta - \dfrac{\theta_0}{2}\right)\right] & \text{при} \quad \theta > \theta_0 \end{cases} \quad (17)$$

где $W_{ph} = P_{ph} t_0$ - энергия импульса света, $\theta_0 = t_0/\tau_d$,

$$U_\omega^{-1} = \frac{q}{\hbar\omega}(1 - R_{ph}) e^{-\kappa x_j} (1 - e^{-\kappa d}), \qquad \mu = \frac{\kappa \tau_d}{e^{\kappa d} - 1} \frac{v_n^2 + v_p^2 e^{\kappa d}}{v_n + v_p} \approx \frac{\tau_d}{t_{np}}.$$

Подстановка (17) в (5) приводит после интегрирования к искомой зависимости тока нагрузки от времени в виде

$$I_L(\theta) = \frac{W_{ph}}{U_\omega t_{np}} \begin{cases} \dfrac{1+\mu}{\theta_0}(\theta - 1 + e^{-\theta}) - \dfrac{\mu \theta^2}{2\theta_0} & \text{при} \quad \theta < \theta_0, \\ (1+\mu)\left(1 - \dfrac{e^{\theta_0} - 1}{\theta_0} e^{-\theta}\right) - \mu\left(\theta - \dfrac{\theta_0}{2}\right) & \text{при} \quad \theta > \theta_0. \end{cases} \quad (18)$$

Результаты расчетов по этим формулам для нескольких значений $t_0$ и $W_{ph}$ приведены на Рис. 4. Нетрудно убедиться, что при $t < t_{n,p}$ выполняется неравенство $(1+\mu)(1 - e^{-\theta}) > \mu\theta$, поэтому функция $I_L(\theta)$ достигает максимального значения

$$\tilde{I} \equiv I_L(\tilde{\theta}) = \frac{W_{ph}}{U_\omega t_{np}}\left[1 - \mu \ln\left(\frac{1+\mu}{\mu} \frac{2}{\theta_0} \operatorname{sh}\frac{\theta_0}{2}\right)\right] \approx \frac{U_0}{R_L} \frac{W_{ph}}{W_0} \quad (19)$$

в момент времени

$$\tilde{\theta} = \ln\left[\frac{1+\mu}{\mu\theta_0}(e^{\theta_0} - 1)\right] > \max(1, \theta_0), \quad (20)$$

где

$$W_0 = \frac{U_0 U_\omega t_{np}}{R_L}\left[1 - \mu \ln\left(\frac{1+\mu}{\mu}\right)\right]^{-1}$$

Последнее приближенное равенство в (19) справедливо, так как вследствие малости $\mu$ второе слагаемое в (19) практически не зависит от $\theta_0$ вплоть до $\theta_0 = 10$ при актуальных значениях остальных параметров. Полученные формулы описывают так называемый линейный режим коммутации, при котором амплитуда тока пропорциональна энергии импульса света и не зависит от напряжения $U_0$. Последняя особенность фотодиодного коммутатора обусловлена независимостью дрейфовых скоростей электронов и дырок от напряженности поля. Так как ток нагрузки не может быть больше $U_0/R_L$, то для реализации линейного режима во всяком случае необходимо выполнение неравенства $W_{ph} < W_0 \approx 2,4$ мкДж. Более точную оценку можно получить, зная распределение электрического поля в освещенной области n-слоя. Для того мы используем вместо уравнения Пуассона закон сохранения полного тока

$$J = j + \varepsilon \frac{\partial E}{\partial t}, \qquad (21)$$

$dJ/dx = 0$. Из (2),(4) и (21) нетрудно получить дифференциальное уравнение

$$\varepsilon \frac{\partial E}{\partial t} = -\frac{\tau_d}{S_d}\frac{dI_L}{dt} + \overline{j} - j, \qquad (22)$$

решение которого имеет вид

$$E(x,t) - E(x,0) \equiv \Delta E(x,t) = \frac{1}{\varepsilon}\left[\int_0^t (\overline{j} - j)dt' - \tau_d \frac{I_L}{S_d}\right], \qquad (23)$$

где $E(x,0) = U_0/d + qN_d(x - d/2)/\varepsilon$ - начальное распределение поля в *n*-слое. Подстановка в (23) соотношений (12)-(15) и (17), (18) приводит после интегрирования к нужным нам[3] распределениям поля в области $v_p t < x < (d - v_n t)$ и значениям в плоскостях $x = 0$, $x = d$:

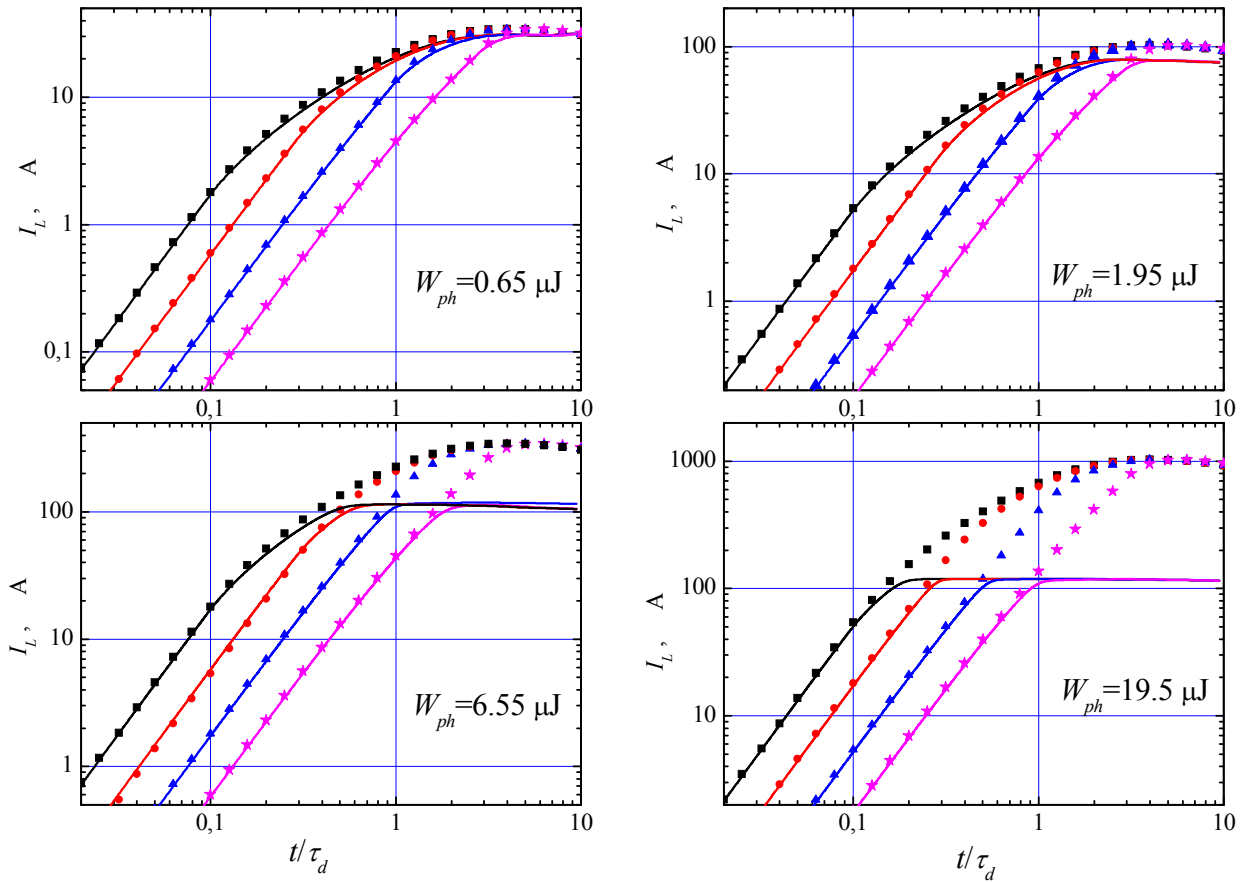

Рис. 4. Ампер-секундные характеристики кремниевого фотодиода в цепи с активной нагрузкой $R_L = 50$ Ом после начала воздействия импульсного освещения при различных энергиях $W_{ph}$ и длительностях $t_0$ импульса: $t_0 = 0.1\tau_d$ (квадраты); $0.3\tau_d$ (кружки); $1.0\tau_d$ (треугольники); $3.0\tau_d$ (звезды). Символы – расчет по формуле (18), линии – результаты численного моделирования.

---

[3] Мы не приводим распределения поля в областях $0 < x < v_p t$ и $(d - v_n t) < x < d$, так они описываются очень громоздкими формулами, а в дальнейшем не используются.

$$\Delta E(x,t) = \frac{W_{ph}R_L}{U_\omega t_{np} d} \begin{cases} \dfrac{\theta^2}{2\theta_0} F(x,t) - \dfrac{S_0}{S_d}(1+\mu)\dfrac{\theta-1+e^{-\theta}}{\theta_0} & \text{при } t < t_0 \\ \left(\theta - \dfrac{\theta_0}{2}\right) F(x,t) - \dfrac{S_0}{S_d}(1+\mu)\left(1 - \dfrac{e^{\theta_0}-1}{\theta_0}e^{-\theta}\right) - \mu\dfrac{\theta}{2}(\theta - \theta_0) & \text{при } t > t_0 \end{cases} \quad (24)$$

$$F(x,t) = 1 + \mu\frac{S_0}{S_d} - \frac{v_n H(d - v_n t - x) + v_p H(x - v_p t)}{v_n + v_p} \frac{\kappa d e^{\kappa x}}{e^{\kappa d} - 1},$$

где $H(x)$ - ступенчатая функция Хевисайда. Пример таких распределений поля приведен на Рис. **5**, а зависимости напряженности поля в «особых» точках от времени изображены на Рис. 6. Резкое уменьшение напряженности поля при удалении от плоскостей $x=0$ и $x=d$ обусловлено наличием нескомпенсированных объемных зарядов электронов и дырок плотностью порядка $qG_0\min(t,t_0)$ в расширяющихся за счет дрейфа областях $0<x<v_p t$ и $(d-v_n t)<x<d$ (см. Рис. 3). Вследствие зависимости скорости генерации пар (11) от координаты в центральной области $v_p t < x < (d - v_n t)$ возникает и увеличивается со временем отрицательный[4] объемный заряд (см. Рис. 3). Плотность этого заряда, пропорциональная $P_{ph}\kappa^2$, относительно невелика вследствие малости $\kappa$. Но при достаточно большой интенсивности света она со временем превосходит заряд доноров $qN_d$ и минимум напряженности поля «перескакивает» в плоскость $x = (d - v_n t)$, как это изображено на Рис. 5,6. Поэтому первое из условий применимости изложенной выше теории имеет вид

$$\min\left[E(v_p t, t), E(d - v_n t, t)\right] > E_s. \quad (25)$$

Здесь следует отметить, что когда нарушается условие $\max\left[E(v_p t, t), E(d - v_n t, t)\right] > E_s$ (это происходит не всегда – см. Рис. 6), то наступает заключительная стадия процесса, рассмотренная в работе [9]. Второе условие применимости теории имеет вид

$$\max(E_0, E_d) < E_b\left[G_0 \min(t, t_0)\right]. \quad (26)$$

Оно связано с увеличением напряженностей поля $E_0(t)$ и $E_d(t)$ на границах *n*-слоя (см. Рис. 6), которые могут превысить пробивное значение $E_b$ резких переходов с концентрацией заряженных примесей в слаболегированных областях порядка $G_0 \min(t, t_0)$ и тогда начинается интенсивное лавинное размножение (в подобных случаях иногда используется термин «динамический лавинный пробой», см., например, [14]), которое не учитывает наша теория.

### 4. Численное моделирование процесса коммутации.

Для описания особенностей процесса коммутации при нарушении условий (25),(26) необходимо решать систему взаимосвязанных нелинейных уравнений непрерывности, Пуассона и внешней цепи, что можно сделать только численными методами. Для этого была использована программа «Исследование» [16]. Основные результаты состоят в следующем.

1) Зависимость амплитуды $I_M$ импульса тока нагрузки, полученной путем численного моделирования, от параметров фотодиода и режима коммутации во всех случаях может быть с достаточной точностью аппроксимирована функцией

---
[4] При противоположном направлении света этот заряд положителен и, увеличивая положительную производную $\partial E/\partial x$, приводит к более раннему нарушению условия $E > E_s$.

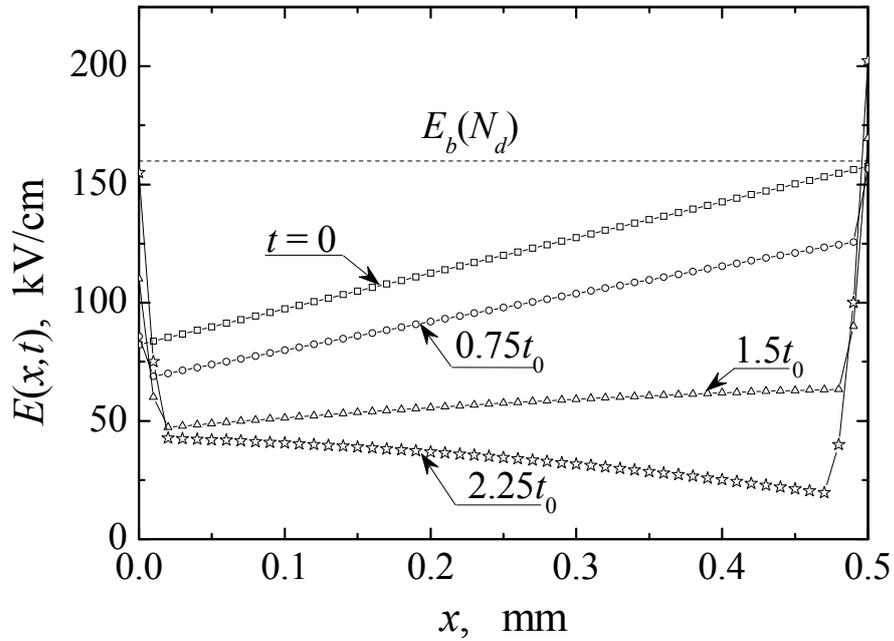

Рис. 5. Распределения электрического поля в *n*-слое фотодиода в моменты времени $t = 0$ (квадраты), $t = 0.75t_0$ (кружки), $t = 1.5t_0$ (треугольники), $t = 2.25t_0$ (звезды) после начала освещения импульсом света с энергией $W_{ph} = 1.95$ мкДж и длительностью $t_0 = 100$ пс. Пунктиром указана пробивная напряженность поле $E_b$ резких кремниевых переходов с концентрацией доноров $N_d = 5 \cdot 10^{14}$ см$^{-3}$ [15],

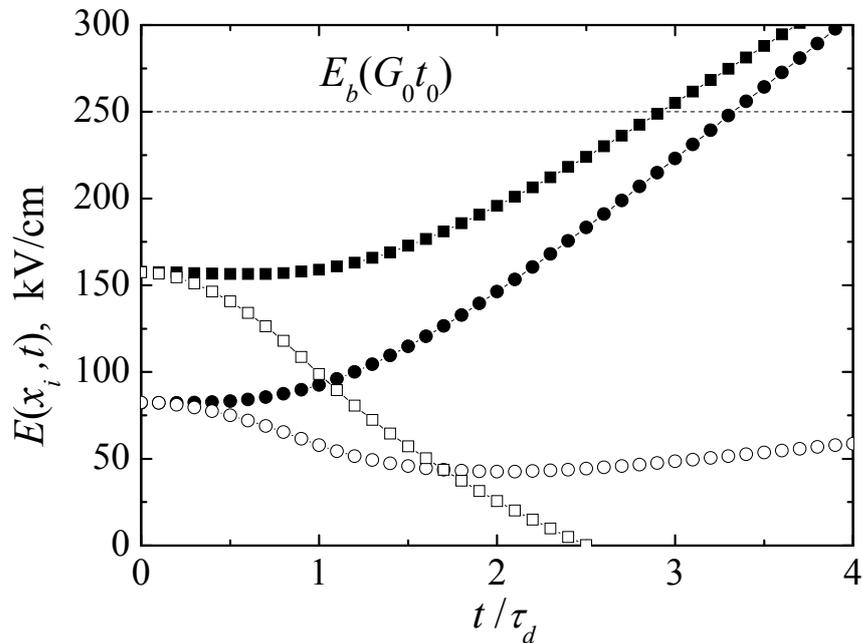

Рис. 6. Зависимости напряженности поля в «особых» точках $x = 0$ (темные кружки), $x = d$ (темные квадраты), $x = v_p t$ (светлые кружки) и $x = d - v_n t$ (светлые квадраты) *n*-слоя от времени после начала освещения импульсом света с энергией $W_{ph} = 1.95$ мкДж и длительностью $t_0 = 100$ пс. Пунктиром указана пробивная напряженность поле $E_b$ резких кремниевых переходов с концентрацией заряженных примесей в слаболегированных областях $G_0 t_0 = 5 \cdot 10^{14}$ см$^{-3}$ [15], соответствующей этому режиму коммутации.

$$I_M \approx \frac{U_0}{R_L} \text{th}\left(\frac{W_{ph}}{W_0}\right), \tag{27}$$

что подтверждается данными, приведенными на Рис. 7. Формула (27) дает правильные предельное значение $I_M = U_0/R_L$ при $W_{ph} > 2W_0$ (когда эффективное сопротивление диода много меньше $R_L$) и значение $I_M = \tilde{I}$, определяемое из (19), при $W_{ph} < W_0/2$. Разумеется, эти необходимые свойства аппроксимаци можно получить не только при использовании в (27) функции $\text{th}(x)$, однако по непонятным причинам она обеспечивает гораздо большую точность в промежуточном интервале значений $W_{ph}$, чем другие простые функции, например, $x/(1+x)$ (сравни с (1)) или $\frac{2}{\pi}\text{arctg}\left(\frac{\pi}{2}x\right)$.

2) Ампер-секундные характеристики, полученные путем численного моделирования и рассчитанные по формулам (18) практически совпадают, пока $I_L(t) \leq 0.9 I_M$ (см. Рис. 4), то есть фактически до завершения процесса коммутации. Это совпадение является следствием того, что, с оной стороны, при актуальных значениях параметров раньше всего нарушается условие (25) применимости теории, а с другой стороны, выполняется приближенное соотношение $E_s < 0.1 E_b(N_d)$. Иными словами, теория фактически применима до тех пор, пока напряжение на фотодиоде больше $E_s d \approx 0.1 U_0$.

3) Результаты численного решения уравнения $I_L(t_{0.9}) = 0.9 I_M$ (определяющего «инженерное» время коммутации $t_{0.9}$) с использованием формул (18) и аппроксимации (27) приведены на Рис. 8. Простые приближенные формулы нетрудно получить в предельных случаях. При малых $W_{ph}$, когда $t_{0.9} \geq t_0$ и $R_d \gg R_L$, время коммутации

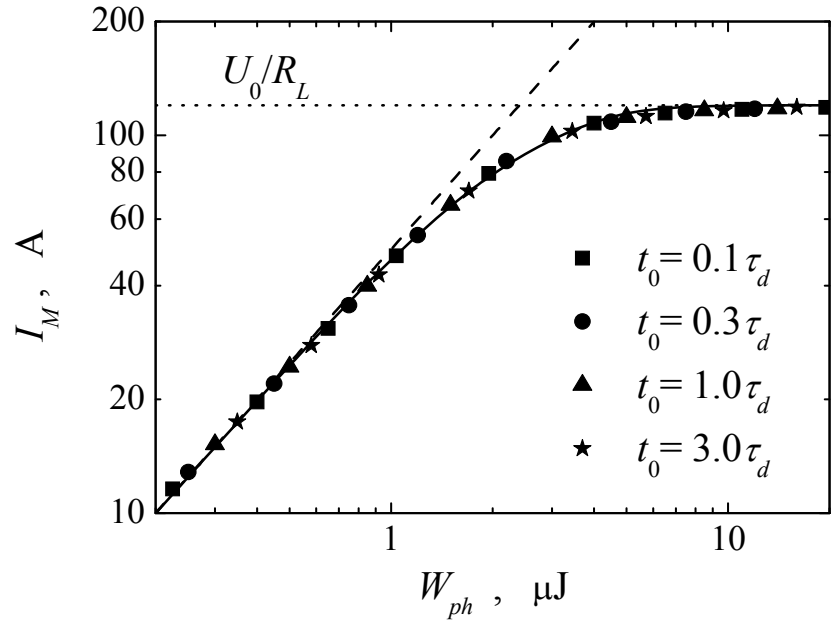

Рис. 7. Зависимость максимального тока нагрузки от энергии импульса света с различной длительностью $t_0$. Символы – результаты численного моделирования, штриховая линия – расчет по формуле (19), сплошная линия – аппроксимация (27) со значением $W_0 = 2.4$ мкДж.

$$t_{0.9} \approx \tau_d \ln \frac{e^{\theta_0}-1}{\theta_0\left(1-0.9\dfrac{U_0 U_\omega t_{np}}{W_{ph} R_L}\operatorname{th}\dfrac{W_{ph}}{W_0}\right)} \approx \tau_d \ln \frac{e^{\theta_0}-1}{\theta_0\left[1-0.9(1+\mu\ln\mu)\right]} \quad (28)$$

и практически не зависит ни от $P_{ph}$, ни от $U_0$. При больших $W_{ph}$, когда $t_{0.9} \leq t_0 < t_d$ и $R_d \ll R_L$,

$$t_{0.9} \approx \sqrt{1.8\tau_d t_{np}\frac{U_0 U_\omega}{P_{ph} R_L}} = \sqrt{1.8\frac{\varepsilon S_d}{v_n+v_p}\frac{U_0 U_\omega}{P_{ph}}} \quad (29)$$

и зависит только от отношения $U_0/P_{ph}$.

Полученные результаты указывают на возможность создания конструктивно простых оптоэлектронных коммутаторов, способных формировать импульсы напряжения от единиц до десятков киловольт с фронтом от десятков до сотен пикосекунд на 50-омной активной нагрузке. Для управления коммутатором можно использовать импульсы света с длительностью порядка десятков пикосекунд и энергией, примерно в 1000 раз меньше, чем а) излучается многими современными коммерческими Nd:YAG лазерами с длиной волны 1,064 мкм, почти идеально подходящей для кремниевых фотодиодов и б) необходима для работы коммутаторов на основе фотосопротивлений [17], использованных при создании сверхширокополосного радара [18]. Развитая в работе теория может служить основой для проектирования подобных коммутаторов с оптимальным сочетанием параметров.

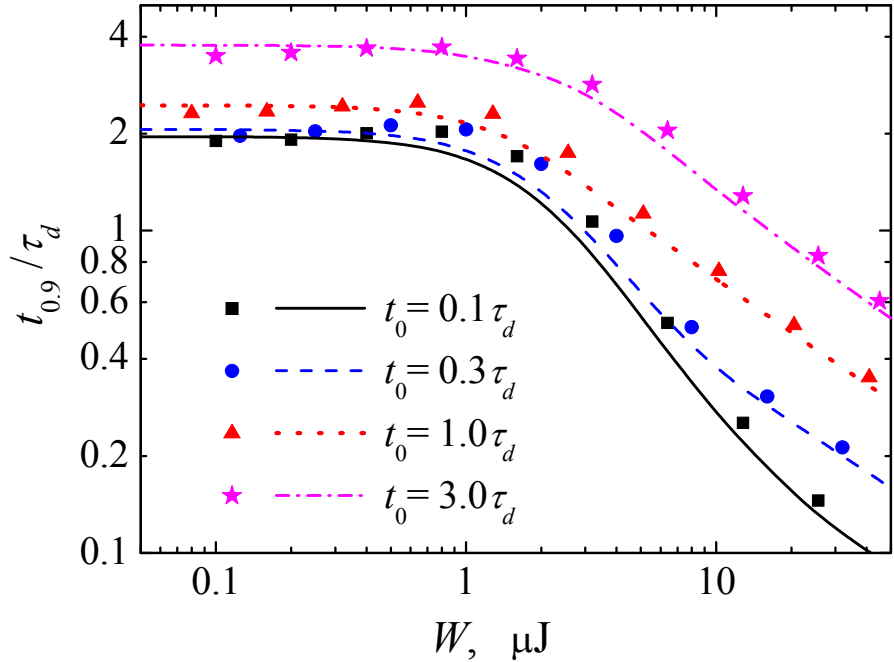

Рис. 8. Зависимости времени коммутации $t_{0.9}$ от энергии импульса света с различной длительностью $t_0$. Символы – результаты численного моделирования, линии – расчет по формуле (28).